\documentclass[showpacs,twocolumn,prl]{revtex4}

\usepackage{amssymb}
\usepackage{amsmath}
\usepackage{epsfig}

\newcommand{\Df}{\Delta F}
\newcommand{\DfPAP}{\Delta F_{P,AP}}
\newcommand{\Hrep}{{\cal H}_{\mathrm{rep}}}

\newcommand{\ab}{_{\alpha\beta}}

\renewcommand{\aa}{_{\alpha\alpha}}

\begin{document}

\title{Interface free energies in $p$-spin glass models}
\author{M. A. Moore}
\affiliation{School of Physics and Astronomy, University of
Manchester, Manchester M13 9PL, UK}

\date{\today}

\begin{abstract}
The  replica method  has been  used  to calculate  the interface  free
energy  associated  with the  change  from  periodic to  anti-periodic
boundary conditions in finite-dimensional $p$-spin glass models in the
phase which at mean-field level has one-step replica symmetry breaking
(1RSB).  In  any  finite   dimension  the  interface  free  energy  is
exponentially small  for a large  system. This result implies  that in
finite dimensions, the  1RSB state does not exist,  as it is destroyed
by thermal excitation of arbitrarily large droplets.  The implications
of this for the theory of structural glasses are discussed.

\end{abstract}

\maketitle

The use of $p$-spin glass  models as models for understanding structural glasses was
 pioneered in  the work of  Kirkpatrick, Thirumalai and  Wolynes 
 \cite{KTW}.  They  found that for the  infinite-range version 
 of these  models, for  which mean-field theory  is exact,  that there
 were two important temperatures.  At a temperature $T_d$ there exists
 a   dynamical  transition  below   which  an   ergodicity  transition
 occurs. The dynamical equations which arise are very similar to those
 of  the  mode-coupling  theory  of  liquids \cite{MCT}.  At  a  lower
 temperature $T_c$ there is a first-order (discontinuous) transition to a
 phase with a 1RSB order parameter. As $T\rightarrow T_c$ the entropy goes to zero, so this
transition  is identified with the Kauzmann glass transition $T_K$
 \cite{Kauzmann}, the temperature at which the extrapolated entropy of
 the supercooled liquid apparently falls to that of the crystalline phase.

 It has been realized for many years
 that outside the mean-field limit there will be no genuine dynamical transition at 
$T_d$. This is because the metastable states which trap the system
 dynamically are unstable  because of activation processes over the finite 
free energy barriers which exist outside the mean-field limit. However, the 
apparent divergence of the viscosity in fragile glasses at some non-zero
 temperature, such as that in the Vogel-Fulcher formula \cite{VK}
\begin{equation}
\log \eta =\log \eta_0 +\frac{\Delta}{T-T_{VF}},
\label{VFE}
\end{equation}
has encouraged  the belief  that there might  be a  real thermodynamic
glass  transition  near  $T_{VF}$.   The observation  that  the  ratio
$T_K/T_{VF}$ lies  beteen 0.9-1.1 for  many glass formers  whose $T_K$
ranges from 50 K to 1000  K \cite{Angell} is further evidence for this
possibility. However, we have recently argued that there is no genuine
glass transition  \cite{MY} and that  $T_K$ and $T_{VF}$ are  just \lq
crossover temperatures'.  The  Ising spin glass in a  field seems to provide 
a satisfactory analogy for supercooled  liquids near $T_{VF}$.

We  shall here  strengthen these  arguments by  showing that  the 1RSB
  state simply does not exist  in the finite dimensional $p$-spin 
  glass model! The  reasons are similar to those why  there is also no
  real dynamical transition $T_d$ in finite dimensional systems. There
  is no  transition at  $T_c$ to a  state with one-step replica  symmetry breaking
  because that state is destabilized by thermal fluctuations within it
  of  droplets of  arbitrarily large  size  which only  cost a  finite
  amount  of  free  energy to  produce.  When  $p$  is even,   the
  Hamiltonian  is left  unchanged by  flipping  the signs  of all  the
  spins, and  the destablizing droplets  are simple to  identify: They
  are droplets of  the time-reversed state. Our basic  task then is to
  calculate the interface free energy for droplets.

There have  recently been other  attempts to calculate  interface free
 energies in  $p$-spin models  \cite{Franz, Dzero}. Their  results are
 very different  to those reported here, essentially  because they are
 studying a  different interface free  energy -- that  appropriate for
 non-equilibrium  effects  below  $T_d$.   The interface  free  energy
 studied here is the one  commonly studied in Ising spin glasses which
 involves changing  the boundary  conditions along one  direction (the
 $z$ direction) from periodic to anti-periodic \cite{AMY}. That is, we
 study   $\delta  F=\sqrt{\overline{\DfPAP^2}}$   (here  and   in  the
 following,  the  overbar means  averaging  over bond  configurations)
 where $\DfPAP = F_{P} - F_{AP}$,  and $F_P$ and $F_{AP}$ are the free
 energies   with  periodic   and  anti-periodic   boundary  conditions
 respectively.  Anti-periodic  boundary conditions can  be realized by
 reversing  the sign of  the bonds  crossing a  plane whose  normal is
 parallel  to  the given  direction.   It  will  be shown  below  that
 $\overline{\DfPAP}= 0$.   If the system is  of length $L$  in the $z$
 direction,  we  find  that  the  interface  free  energy  $\delta  F$
 decreases  to  zero as  $\sim  \exp(-L/2\xi)$,  provided $L\gg  \xi$.
 $\xi$ is the  correlation length in the 1RSB  state.  This state will
 therefore  be  destabilized  by   the  thermal  excitation  of  large
 droplets,  since the  free  energy  cost of  creating  them --  their
 interface free energy-- can be made arbitrarily small.

There exists  simulational evidence as to the validity of the
conclusion that  the 1RSB state does  not exist in  the 10-state Potts
spin  glass \cite{Potts}.  In the  infinite-range limit  when  all the
Potts  spins are  coupled  by random  interactions, mean-field  theory
becomes exact  and is  the same  as that for  the $p$-spin  model. The
simulations  of  it were  consistent  with  the  existence of  both  a
dynamical  transition $T_d$  and a  transition to  a 1RSB  state below
$T_c$.  However, in three  dimensions using a Gaussian distribution of
the couplings, there was no  evidence of a dynamical transition $T_d$,
nor of a transition to a 1RSB state.

 The complete  absence in three  dimensions of the  dynamical behavior
found at  mean-field level  in the Potts  glass is  perhaps surprising
given  that  in  supercooled  liquids  there is  ample  evidence  that
mode-coupling works  well. Furthermore we  have recently been  able to
{\em derive}  the replicated form of the  $p$-spin functional starting
directly from the liquid  state \cite{MY}. However, this derivation of
the  $p$-spin  functional  provides  the  answer to  the  puzzle:  Its
variables  are  quantities  which  are  not directly  related  to  the
densities  whose correlations functions  are studied  in mode-coupling
theory.   The derived  replicated p-spin  functional is  useful though
when studying  whether there might  be a genuine glass  transition. It
can  be  further  transformed   to  the  replicated  functional  which
describes an Ising spin glass in a field \cite{MY, MD}. Then the issue
of whether there is or is not a genuine thermodynamic glass transition
is the same as whether there is a transition of an Ising spin glass in
a magnetic  field. But whether  there is or  is not a transition  to a
glass state, that state cannot be the 1RSB state as in the mean-field
$p$-spin glass model, since such a state does not exist in finite dimensions.

Numerical  studies of  $p$-spin  models outside  the mean-field  limit
 \cite{PPR} use Hamiltonians which are of the following form, which is
 appropriate for the  case $p=3$ \cite{PPR}. Each site  is occupied by
 two Ising spins, $\sigma_i$ and $\tau_i$, and the Hamiltonian is
\begin{eqnarray}
{\mathcal H}(\{\sigma\},  \{\tau\})= &-&\sum_{<ij>}\left( J_{ij}^{(1)}
\sigma_i\tau_i\sigma_j        +J_{ij}^{(2)}       \sigma_i\tau_i\tau_j
\right.  \nonumber\\  &&{}+\left.J_{ij}^{(3)}\sigma_i\sigma_j\tau_j  +
J_{ij}^{(4)}\tau_i\sigma_j\tau_j\right),
\label{H3}
\end{eqnarray}
where   ${<ij>}$  are  nearest-neighbor   pairs,  and   the  couplings
 $J_{ij}^{(n)}$ are chosen  independently from a Gaussian distribution
 with zero  mean and width $J$.  Note  that when the signs  of all the
 spins are  reversed, the sign of the  Hamiltonian changes, indicating
 the violation of  time reversal symmetry.  For models  which have $p$
 even, such as  that for $p=4$ in Ref.  \cite{FP}, the Hamiltonian does
 not change sign when all the spins are flipped.

Starting from a $p$-spin Hamiltonian such as that in Eq. (\ref{H3}) it
is possible to  obtain the \lq generic' form  of the replicated p-spin
functional \cite{MD,  Ferrero, Dzero,  MYunpub} by averaging  over the
couplings:
\begin{multline}
\beta \Hrep  = \int  d^d x \,  \left[ {t \over  4} \sum_{\alpha,\beta}
q\ab^2 + {1 \over 4} \sum_{\alpha,\beta} (\vec{\nabla} q\ab)^2 \right.
\label{Hrep} \\
\left.     -{w_1     \over    6}    \sum_{\alpha,\beta,\gamma}    q\ab
q_{\beta\gamma}q_{\gamma\alpha}  -{w_2  \over  6}  \sum_{\alpha,\beta}
q\ab^3 +{y \over 24} \sum_{\alpha,\beta} q\ab^4 \right] ,
\end{multline}
where    $q_{\alpha\beta}$     is    a    symmetric     matrix    with
 $q_{\alpha\alpha}=0$.   The coefficients  $t, w_1,  w_2$ and  $y$ are
 arbitrary positive  parameters. This generic functional  has the same
 relationship to the finite dimensional $p$-spin glass models as (say)
 the Landau-Ginzburg functional has  to the Ising model. It represents
 a  $p$-spin model  with odd  values  of $p$  as the  cubic term  with
 coefficient $w_2$  corresponds to  a term which  breaks time-reversal
 invariance. (Terms in which a  replica index appears an odd number of
 times  always break  time-reversal invariance).   The odd  $p$ models
 would seem  to be  the relevant ones  for applications  to structural
 glasses which have no identifiable symmetries. The replica indices go
 from 1 to $n$, and $q\aa=0$.

In order  to calculate interface  free energies between  systems which
differ  as to  whether they  have periodic  or  anti-periodic boundary
conditions, we will follow the procedure used in Ref. \cite{AMY}. This
requires one to replicate the system with periodic boundary conditions
$n$ times  and the system  with anti-periodic boundary  conditions $m$
times, where as usual  we will take $n$ and $m$ to  zero at the end of
the calculation.
 
 Expanding the replicated partition function in powers of $m$ and $n$,
we have
\begin{multline}
-\ln \overline{ Z_P^n Z_{AP}^m } =  (n + m) \, \beta \overline{F} \\ -
{(n+m)^2\over 2}\, \beta^2 \overline{\Df^2}  + {nm \over 2} \, \beta^2
\overline{\DfPAP^2} + \cdots ,
\label{lnZnZm}
\end{multline}
where  $  \overline{\Df^2}  =  \overline{F_P^2} -  \overline{F_P}^2  =
\overline{F_{AP}^2}  -  \overline{F_{AP}}^2  $  is the  (mean  square)
sample-to-sample  fluctuation of the  free energy,  the same  for both
sets  of  boundary  conditions   $P$  or  $AP$,  and  $\overline{F}  =
\overline{F_P} = \overline{F_{AP}}$.  Thus $\overline{\DfPAP}= 0$.  To
find the variance of the interface free energy, $\overline{\DfPAP^2}$,
we expand out $\ln \overline{ Z_P^n Z_{AP}^m }$ to second order in the
numbers of  replicas, $n$ and  $m$, separate out the  pieces involving
the \textit{total}  number of replicas  $n+m$, and take  the remaining
piece, which is proportional to $n m$.

The  replica indices  in  the generalization  of  Eq. (\ref{Hrep})  go
 $\alpha,\beta,\gamma=1,  2, \cdots,  n, n+1,\cdots,  n+m$.  The order
 parameter $q$ divides naturally into blocks of size $n$ and $m$. From
 now  on, Greek indices  will label  the first  block, Roman  ones the
 second block, so, for example, $q_{\alpha i}$, means $\alpha\in[1,n]$
 and  $i\in[n+1,n+m]$,  and refers  to  the  respective  entry in  the
 off-diagonal, or mixed, sector.

The boundary condition which results  from the flipping the sign of of
the bonds in the plane perpendicular  to the $z$-axis is that $q$ must
be periodic in  the Greek and Roman sectors,  and anti-periodic in the
mixed sectors:
\begin{equation}
\begin{split}
q_{\alpha\beta}(z) &= q_{\alpha\beta}(z+L) \\ q_{ij}(z) &= q_{ij}(z+L)
  \\ q_{\alpha i}(z) &= -q_{\alpha i}(z+L).
\end{split}
\label{bcAP}
\end{equation}
In the directions perpendicular to the $z$-axis, which we will take to
 be all of length $M$, the system is periodic.
 
 At   mean-field  level,   the  \textit{stable}   solution   for  $\ln
\overline{Z_P^n Z_{AP}^m}$ is given by
\begin{align}
-\ln \overline{Z_P^n Z_{AP}^m} &= \beta\Hrep\{q^{\text{SP}}\},
\end{align}
where $q^{SP}$ denotes the solution of the Euler-Lagrange equations:
\begin{equation}
\begin{split}
\frac{d^2q\ab}{dz^2}&=tq\ab-w_1\sum_{\gamma=1}^nq_{\alpha\gamma}q_{\gamma\beta}
- w_1\sum_{i=1}^{m}q_{\alpha  i}q_{i\beta}\\ &-w_2q\ab^2+yq\ab^3/3, \\
\frac{d^2q_{ij}}{dz^2}&=tq_{ij}-w_1\sum_{k=1}^mq_{ik}q_{kj}
-w_1\sum_{\gamma=1}^{n}q_{i\gamma}q_{\gamma           j}\\           &
-w_2q_{ij}^2+yq_{ij}^3/3,\\  \frac{d^2q_{\alpha  i}}{dz^2}&=tq_{\alpha
i}-       w_1\sum_{\gamma=1}^n       q_{\alpha       i}q_{\gamma\beta}
-w_1\sum_{k=1}^mq_{\alpha  k}q_{k i}\\  &-w_2q_{\alpha i}^2+yq_{\alpha
i}^3/3.
\end{split}
\end{equation}
Notice that there is a solution of these equations which satisfies the
boundary conditions  in which the mixed overlaps  $q_{\alpha i}=0$ and
where in  the all Greek or  all Roman sectors $q\ab$  and $q_{ij}$, the
solutions are $z$ independent. We  believe that this is the physically
relevant solution as  it is stable. Thus at  mean-field level the free
energy  of the $n+m$  replicated system  with the  boundary conditions
breaks  up into  the  sum of  the free  energies  of the  $n$ and  $m$
replicated systems  without boundary conditions, so that  there can be
no term  of order  $nm$ in Eq.  (\ref{lnZnZm}), so that  the interface
energy vanishes to this order.

 In  the  high-temperature  phase  the solution  to  the  stationarity
equations is $q\ab=q_{ij} =q_{\alpha i} =0$. A spatially constant 1RSB
solution  appears  discontinuously  at a  temperature  $t_K=2w^2/(3y)$
where to  simplify the algebra  the following notational  changes have
been made: $w_1=1$,  $w_2=1+w$ with $w>0$. (For the  specific model of
Eq. (\ref{H3}) $w$ is actually negative \cite{MYunpub}). There is, for
temperatures  $t\leq t_K$,  a 1RSB  solution in  which $q\ab$  is zero
everywhere except in $n/\tilde{m}$ boxes of size $\tilde{m}$ along the
leading diagonal,  where it takes the value  $q_1$. $q_{ij}$ similarly
is   zero   everywhere  except   in   $m/\tilde{m}$   boxes  of   size
$\tilde{m}$.   $q_{\alpha   i}=0$.   For   a   review   of  1RSB   see
Ref.  \cite{Cavagna}. Right  at $t_K$,  on letting  $n,m$ go  to zero,
$q_1=2w/y$   and  $\tilde{m}=1$.   For  temperatures   $t<t_K$,  $q_1$
increases and $\tilde{m}$ decreases.

To get  the leading non-vanishing  contribution to the  interface free
 energy we need  to go to one-loop order. The  first correction is due
 to Gaussian fluctuations around the stationary 1RSB solution:
\begin{align}
-\ln \overline{Z_P^n Z_{AP}^m}  &= \beta\Hrep\{q^{\text{SP}}\} + \frac
  12 \sum_k I(k^2),
\end{align}
where
\begin{align}
    \label{Idef}
    I(k^2) &= \sum_{\mu}d_{\mu}\ln(k^2+\lambda_{\mu}).
\end{align}
$\vec{k}$ is  a $d$-dimensional  wave vector, $\lambda_{\mu}$  are the
eigenvalues  of the Hessian,  evaluated at  the stationary  point, and
their degeneracies are $d_{\mu}$. The eigenvalues and degeneracies are
the  same as for  a system  of size  $n+m$ without  boundary condition
changes (because the stationary point is the same), only the nature of
the  $k$-vectors changes  for  the terms  involving eigenvalues  whose
corresponding eigenvectors  $f$ are  nonzero exclusively in  the mixed
sector (i.e.\  $f\ab=f_{ij}=0$): the wave vectors have  to respect the
imposed        boundary        conditions,        which        implies
$\vec{k}=(2n_1\pi/M,\dots,2n_{d-1}\pi/M,(2n_d+1)\pi/L)$           (with
$n_i\in\mathbb{Z}$)    in   the   mixed    sector   as    opposed   to
$\vec{k}=(2n_1\pi/M,\dots,2n_{d-1}\pi/M,2n_d\pi/L)$  in  the Greek  or
Roman sectors.

While it  is possible to  compute the eigenvalues  $\lambda_{\mu}$ and
their  degeneracies  $d_{\mu}$  it   is  much  easier  to  follow  the
procedures  of  Ref.   \cite{AspelmeierMoore02,  AMY}.   One  computes
$\partial I/\partial(k^2)$ rather than  $I$ itself. This helps because
it is possible to express the terms in this derivative of order $n^2$,
$m^2$ and $nm$  directly in terms of a single  propagator of the mixed
sector calculated for the system without boundary condition changes:
 \begin{align}
\label{Imixed}
\frac{\partial I_{AP}}{\partial (k^2)} &= nm G_{\alpha i, \alpha i}
\end{align}
where the  prefactor $nm$ reflects  the number of eigenvectors  in the
 mixed sector. The  subscript AP indicates that its  argument $k$ must
 have values appropriate to the mixed sector. Defining the integral
\begin{equation}
J(k):=\int d(k^2)G_{\alpha i, \alpha i}
\end{equation}
and    obtaining    the    constant    of   integration    via    Ref.
\cite{AspelmeierMoore02}  enables  one to  write  the  terms of  order
$n^2$, $m^2$ and $nm$ in $I$ as
\begin{align}
 &  \frac{n^2+m^2}{2}J_{P}(k^2)  +  nm  J_{AP}(k^2)  \nonumber  \\  &=
\frac{(n+m)^2}{2}J_{P}(k^2) + nm(J_{AP}(k^2)-J_{P}(k^2)).  \nonumber
\end{align}
 The subscripts $P$ and $AP$ on $J$ mean that $J$ must be taken as $0$
when  the argument  is  not of  the  required type,  i.e. periodic  or
anti-periodic.

Comparison with Eq.~\eqref{lnZnZm} shows
\begin{multline}
\beta^2\overline{\DfPAP^2}    =    \left({\sum}_{AP}-{\sum}_{P}\right)
  J(k^2)       =       \\       \sum_l       \sum_{r=-\infty}^{\infty}
  \left(J(l^2+\frac{(2r+1)^2\pi^2}{L^2})                              -
  J(l^2+\frac{(2r)^2\pi^2}{L^2})\right)
\label{ienergy}
\end{multline}
where the  subscripts on the sums  indicate the nature  of the allowed
$k$-vectors, as made explicit in the second part of the equation where
the $z$  component of the $k$-vector  has been split  off, leaving the
$d-1$-dimensional wave vector $l$. The  sum over the $z$ component can
be  extended  to  $\pm\infty$,  introducing only  exponentially  small
errors for large $L$.

 The calculation  of $G_{\alpha i,\alpha i} $  is readily accomplished
using the  methods of Ref.  \cite{BM}. Right at the  transition $t_K$,
the  propagator  is  very  simple  because  $\tilde{m}=1$:  $G_{\alpha
i,\alpha  i}=  1/(k^2+t)$,  so  $J(k^2)=\ln  (k^2+t)$.  We  shall  for
simplicity  work in  the limit  where the  sums over  the  first $d-1$
directions  can be  converted to  integrals (renaming  the integration
variable q). Then
\begin{multline}
\beta^2           \overline{\DfPAP^2}           =          M^{d-1}\int
 \frac{d^{d-1}q}{(2\pi)^{d-1}}  \sum_{r=-\infty}^{\infty}\\  \left(\ln
 (q^2+\frac{(2r+1)^2\pi^2}{L^2}                +t)-                \ln
 (q^2+\frac{(2r)^2\pi^2}{L^2}+t)\right).  \nonumber
\end{multline}
The sum  over $r$  can be done  analytically when the  right-hand side
becomes
\begin{multline}     
 2\frac{S_{d-1}}{(2\pi)^{d-1}}M^{d-1}\int_0^{\infty}\,dq        q^{d-2}
 \ln\coth[\sqrt{q^2+t}L/2].  \nonumber
\end{multline}
$S_d$ is  the surface  area of a  $d$-dimensional unit sphere.  In the
limit  when  $L\gg \xi$,  where  $\xi=1/\sqrt{t}$  is the  correlation
length, the integral is simple to evaluate:
\begin{multline}
\label{largeL}
\beta^2     \overline{\DfPAP^2}     \approx    2^{\frac{d+1}{2}}\Gamma
\left(\frac{d-1}{2}\right)   \frac{S_{d-1}}{(2\pi)^{d-1}}   \\  \times
\left(\frac{M}{L}\right)^{d-1}
\left(\frac{L}{\xi}\right)^{\frac{d-1}{2}}
\exp\left(-\frac{L}{\xi}\right).
\end{multline}
This result is valid in the limits $M \gg L \gg \xi$. By going back to
 Eq. (\ref{ienergy}), however, any desired aspect ratio can be studied
 for any value of $L/\xi$.

While we have produced an explicit expression for the interface energy
$\delta F$ at  one particular temperature, $t_K$, at  which the system
jumps  into the  1RSB  state, the  same  methods can  be  used at  any
temperature. The origin of the  exponentially small free energy is the
presence of  positive non-zero eigenvalues in the  Hessian matrix. The
stationary  point  needs to  be  marginally  stable  in order  to  get
interface free energies  of the form $\delta F  \sim L^{\theta}$, with
$\theta  >  0$,  which  is  what  happens for  the  Ising  spin  glass
\cite{AMY}.   In  both  the  temperature  regions  \textit{above}  and
\textit{below} $t_K$,  the eigenvalues  are positive so  the interface
free energy  in $p$-spin models  is at all  temperatures exponentially
small.  This  conclusion will  also be true  of the 1FRSB  state which
exists for some generalizations of the $p$-spin model, where the order
parameter  $q(x)$ has  both a  jump,  characteristic of  1RSB, but  in
addition varies with  $x$, like in full replica  symmetry breaking. It
has also only non-zero Hessian eigenvalues \cite{CL}.

The methods used here can also be used to calculate the interface free
energy  appropriate  to  the   dynamical  questions  studied  in  Ref.
\cite{Franz, Dzero}.   To locate the dynamical  transition one imposes
the  condition  that the  stationary  point  solution  $q^{SP}$ has  a
Hessian with a null eigenvalue \cite{CL}. This will guarantee that the
associated interface free energy will not be exponentially small.  The
calculation is very similar to that of Ref. \cite{AMY} \cite{MM}.

\begin{acknowledgments}
I should like to thank the Isaac Newton Insitute, Cambridge, UK for 
hospitality during the writing of this paper and G. Biroli, J-P Bouchaud,
 Silvio Franz and other PDS Workshop participants for stimulating discussions.
\end{acknowledgments}

\end{document}